\documentclass[letters,usenatbib]{mnras}

\usepackage{natbib}
\bibpunct{(}{)}{;}{a}{}{,}
\usepackage{graphicx}
\usepackage{txfonts}
\usepackage{xcolor}
\usepackage{multirow}
\usepackage{pst-plot}
\usepackage{breakurl}
\SpecialCoor
\usepackage{hyperref}

\def\xmm{{\em XMM-Newton}}

\def\chan{{\em Chandra}}

\def\xrism{{\em XRISM}}

 \definecolor{arancio}{rgb}{1,0.5,0}
 \definecolor{viola}{rgb}{0.7,0,1}
 \definecolor{verde}{rgb}{0.2,0.7,0.7}
\definecolor{cobalt}{rgb}{0.0, 0.28, 0.67}
\definecolor{airforceblue}{rgb}{0.36, 0.54, 0.66}
\definecolor{ballblue}{rgb}{0.13, 0.67, 0.8}
\definecolor{battleshipgrey}{rgb}{0.52, 0.52, 0.51}
\definecolor{darkgreen}{rgb}{0.0, 0.2, 0.13}

\title[\chan/HETG Doppler velocity measurements from Capella]{A detailed analysis of X-ray emission line velocities of Capella from over 20 years of \chan/HETG spectroscopy}
\author[E. Bozzo et al.]{
E.\ Bozzo$^{1}$\thanks{E-mail: enrico.bozzo@unige.ch},
D. P.\ Huenemoerder$^{2}$,
N. Produit$^{1}$, 
M. Falanga$^{3,4}$,
S. Paltani$^{1}$,
E. Costantini$^{5,6}$
\\
$^{1}$Department of Astronomy, University of Geneva, Chemin d'Ecogia 16, CH-1290 Versoix, Switzerland\\
$^{2}$Massachusetts Institute of Technology, Kavli Institute for Astrophysics and Space Research, 77 Massachusetts Ave., Cambridge, MA 02139, USA \\
$^{3}$International Space Science Institute (ISSI), Hallerstrasse 6, 3012 Bern, Switzerland \\
$^{4}$Physikalisches Institut, University of Bern, Sidlerstrasse 5, 3012 Bern, Switzerland\\
$^{5}$SRON Netherlands Institute for Space Research, Niels Bohrweg 4, 2333 CA Leiden, The Netherlands\\
$^{6}$Anton Pannekoek Institute, University of Amsterdam, Postbus 94249, NL-1090 GE Amsterdam, the Netherlands
}

\date{}

\begin{document}
\label{firstpage}
\pagerange{\pageref{firstpage}--\pageref{lastpage}}
\maketitle

\begin{abstract}
Capella is the brightest chromospherically active binary in the sky, hosting a cooler G8III giant (Aa) and an hotter G1III companion (Ab). The source has been extensively observed in the X-rays in the past decades not only for its astrophysical interest in the field of corona sources, but also for in-flight calibrations of space-based X-ray instruments. 
In 2006, it was demonstrated using \chan/HETG observations that Aa is the main contributor to Capella's X-ray emission, as the centroid energies of the emission lines are Doppler shifted along the orbit of the G8III  giant (an aspect that has to be taken in consideration for calibration activities of X-ray instruments). In this paper, we extend the previous analysis performed in 2006 by re-analyzing in an homogeneous way all \chan/HETG observations performed in the direction of Capella. By doubling the amount of data available, we strengthened the conclusion that Capella Aa is the dominant emitter in soft X-rays. We did not find any evidence of a statistically significant contribution to this emission by the Ab giant. Our findings are discussed also in light of the incoming launch of the \xrism\ mission (spring 2023).  
\end{abstract}


\section{Introduction}
\label{sec:intro}

Capella is a well known binary system, hosting two giant stars in a non-eclipsing orbit
with a measured orbital period of about 104~days. The system geometry (orbital axes, inclination, eccentricity, ...) has 
been accurately determined over the decades, with the most recent results reported by \citet{strassmeier20}. 
The two stars in Capella are classified as giants of spectral types G8III and G1III, commonly 
labeled as Aa and Ab, respectively. These stars are characterized by relatively similar masses of about 2.5~M$_{\odot}$. 
The system is recognized to be the brightest chromospherically active binary in the sky, with the slightly more massive G8III star being 
a core He-burning clump giant and the hotter G1III star a rapidly rotating giant still in the Hertzsprung gap 
\citep[see, e.g.,][for recent reviews]{torres09,weber11,torres15}.
\begin{table*}
\scriptsize
\caption{\chan/HETG observations of Capella considered in this work. Observations 1099-5955 were those already reported by ISH06, while observations 6471-25616 are reported here for the first time. For each observation, we report the ID, the start time in heliocentric Julian day, the effective exposure time, the orbital phase calculated by using the latest available Capella's ephemerides \citep{strassmeier20}, the measured apparent velocity obtained from the \chan\ data, as well as the corrected velocity for the barycentric motion of the Earth around the center of mass of the solar system. The uncertainties indicated for the apparent velocity are given at 3$\sigma$ c.l.}
\label{tab:obs}
\begin{tabular}{ccccccc}
\hline
 &  & Effective &   & Apparent Velocity & Corrected Velocity & Barycenter \\
 & Date Start & exposure & Phase & V$_{app}$ & V$_{bary}$ & Correction  \\
ObsID & (HJD)  & (ksec)   & $\phi$ & (km/s) & (km/s) & (C$_{bary}$ km/s) \\
\hline
1099  & 2451418.827 &  14.6 & 0.400 & -42.1$_{-18.9}^{+18.1}$   & -15.9 &  26.0 \\ 
1235  & 2451419.010 &  14.6 & 0.402 & -29.4$_{-18.1}^{+19.1}$   & -3.1 & 26.0 \\
1100  & 2451419.192 &  14.6 & 0.403 & -25.9$_{-22.0}^{+21.7}$   & 0.6 & 26.1 \\
1236  & 2451419.375 &  14.6 & 0.405 & -19.5$_{-19.1}^{+21.9}$   & 7.0 & 26.1 \\
1101  & 2451419.558 &  14.6 & 0.407 & -24.6$_{-17.1}^{+17.5}$   & 1.9 & 26.1 \\
1237  & 2451419.741 &  14.6 & 0.409 & -19.1$_{-20.0}^{+19.0}$   & 7.5 & 26.1 \\
1103  & 2451445.758 &  40.5 & 0.659 & -9.3$_{-11.3}^{+11.6}$    & 18.0 & 26.9  \\
1318  & 2451447.061 &  26.7 & 0.671 & -10.3$_{-16.1}^{+14.9}$   & 16.3 & 26.8 \\
0057  & 2451607.188 &  28.8 & 0.211 &  61.5$_{-13.6}^{+14.8}$   & 34.6 & -27.3 \\
1010  & 2451952.019 &  29.5 & 0.526 &  42.3$_{-15.6}^{+15.0}$   & 17.7 & -24.2 \\
2583  & 2452394.237 &  27.6 & 0.777 &  58.1$_{-13.1}^{+13.7}$   & 38.8 & -18.8 \\
3674  & 2452910.319 &  28.7 & 0.738 &  2.0$_{-16.8}^{+17.9}$  & 28.1 & 26.7 \\
5040  & 2453259.424 &  28.7 & 0.094 &  14.1$_{-14.4}^{+13.6}$   & 40.6 & 27.3 \\
5955  & 2453458.052 &  28.7 & 0.004 &  73.4$_{-12.8}^{+12.8}$   & 46.9 & -26.5 \\
\hline
\hline
6471 &	2453843.643 &	29.6 &	0.711 &	49.7$_{-12.1}^{+12.8}$   & 26.9 &	22.4 \\
9638 &	2454576.408 &	31.0 &	0.755 &	55.4$_{-10.1}^{+10.7}$   & 34.9 &	21.9 \\
10599 &	2454944.046 &	29.2 &	0.289 &	55.3$_{-12.2}^{+13.0}$   & 28.4 &	21.0 \\
11931 &	2455154.395 &	29.6 &	0.311 &	13.5$_{-13.2}^{+13.2}$   & 24.6 &	-11.6 \\
13089 &	2455531.575 &	29.6 &	0.937 &	33.8$_{-15.5}^{+15.0}$   & 38.8 &	-6.1 \\
14239 &	2455925.124 &	29.5 &	0.721 &	46.2$_{-15.8}^{+15.5}$   & 36.0 &	7.5 \\
16418 &	2456649.930 &	29.5 &	0.688 &	32.1$_{-15.6}^{+16.0}$   & 23.5 &	4.8 \\
17324 &	2456992.652 &	28.7 &	0.983 &	31.0$_{-18.2}^{+17.6}$   & 35.5 &	-6.1 \\
18357 &	2457596.027 &	14.8 &	0.784 &	10.1$_{-24.0}^{+24.1}$   & 27.0  &     -18.2 \\
18364 &	2457596.730 &	14.8 &	0.790 &	12.3$_{-21.3}^{+21.6}$   & 27.9 &	-18.4 \\
21786 &	2458455.466 &   17.8 &	0.046  & 30.0$_{-24.9}^{+29.3}$   & 36.7	& -5.3 \\
22003 &	2458457.772 &	11.8 &	0.068 &	52.4$_{-29.2}^{+31.0}$   & 53.4 &	-4.2 \\
25616 &	2459563.322 &	27.6 &	0.696 &	25.4$_{-21.7}^{+22.0}$   & 24.2 &	0.6 \\
\hline
\end{tabular}
\end{table*}

Capella is a relatively bright nearby X-ray source, achieving a luminosity of about 1.5$\times$10$^{28}$~erg~s$^{-1}$ in the 
0.3-5~keV energy band \citep[at a distance of $\sim$13~pc;][]{singh20}. The widely debated question in the late 90's on 
which of the two giant stars in the binary was contributing more to the X-ray luminosity was first addressed in \citet[][hereafter ISH06]{ishibashi06}. These authors used all 14 observations of Capella that were available at that time with the High Energy Transmission Grating \citep[HETG;][]{hetg} on-board the \chan\ satellite \citep{chandra} to show that the identified X-ray emission lines in the spectrum of the source were 
moving in energy as a consequence of the Doppler effect within the binary. The detected displacements in energy were shown to match quantitatively well those expected along the orbit of Capella's Aa component. 
This proved that the coolest G8III giant was providing the largest contribution to the overall Capella high energy emission, at 
least in the period covered by the \chan\ data (1999 August - 2005 March). ISH06 also tentatively identified a portion of the orbit close to the time of the passage of Aa  through the ascending node (phases 0.002-0.092) where the measured line Doppler velocities were somewhat lower than expected, which could have indicated a significant  contamination by the high energy emission of the secondary Ab component. This was supported by the analogy with the known emission properties of the K0 giant $\beta$~Cet and the Hertzsprung gap G0 giant 31~Com. 

Apart from the astrophysical interest for Capella as a prototypical coronal source, the system has also gained an increasing interest in the past decades due to the richness of emission features in its soft X-ray spectrum (0.3-5~keV) and the long-term stability of its X-ray flux. When combined with the source's relatively high flux, these properties have made Capella a well suited celestial object for in-flight calibrations of X-ray instruments designed for high resolution spectroscopy. Among these, the reflection grating spectrometer (RGS) on-board \xmm\ \citep{rgs} observed often Capella over the years \citep[see, e.g.,][and references therein]{audard01} to perform the periodic calibrations of the instrument wavelength/energy scale (i.e., the relation between the wavelength/energy of the detected photon and its most likely location on the detector), the instrument CCD electronic readout gain (i.e., the conversion factor from the recorded electric charge to the corresponding readout pulse height), and the serial calibrations of the CCD charge transfer inefficiency \citep[CTI;][]{vries}. Similar activities have been performed for both the low energy transmission grating \citep[LETG;][]{letg0} and the HETG on-board \chan.\ Capella observations have been exploited to calibrate the instrument dispersion relations, their resolving powers, their line response functions, and the grating-detector alignments \citep[see, e.g.,][and reference therein]{letg,letg2,letg3,hetg}. A detailed knowledge of the physical properties of the emission lines in Capella's X-ray spectrum, including their centroid energies and the intrinsic variations over time due to Doppler effects within the binary, is thus critical not only to understand the physics of coronal sources but also to assess and monitor along the years the performance of operating X-ray instruments. Calibration data have to take into account and compensate for the physical variability of the lines in order to single out systematics that are intrinsic to the instrument.   

In the coming future, our ability to perform high resolution astrophysical X-ray observations will be boosted by the launch of the \xrism\ mission \citep{xrism1,xrism2}, which launch is expected in spring 2023. The \xrism/Resolve instrument will achieve an energy resolution of $\sim$5~eV across the covered energy band (0.3-12~keV) and provides a much larger effective area compared to the \chan\ and \xmm\ gratings. This opens up the possibility of performing high resolution spectroscopic observations for a number of faint and previously out of reach classes of celestial sources \citep{xrismscience}. As part of its calibration plan, \xrism/Resolve is also aiming at periodic observations of Capella \citep{xrismcal}. 

In this paper, we extend the previous work of ISH06 by analyzing all the available Capella observations performed up to the end of 2022 with the \chan/HETG. Our dataset includes the 14 observations previously reported by ISH06 and other additional 13 observations not yet analyzed to study the Doppler effects of the binary on the centroid energy of the most prominently detected emission lines. We carried out an homogeneous analysis of all data by using the latest version of the \chan\ processing software and calibration files released at the time of writing. We discuss the implications of the Doppler measurements for the calibrations of the incoming \xrism/Resolve instrument.

\section{Data processing and analysis}
\label{sec:data}

In order to ease the comparison with the results reported previously by ISH06, we followed in this paper a similar procedure for the processing and analysis of our 
extended \chan/HETG dataset. All considered observations are listed in Table~\ref{tab:obs}. These correspond to all available data collected in the direction of Capella between the beginning of \chan\ science operations in 1999 and the end of 2022. We selected only the observations where the HETG was used in combination with the ACIS-S. This combination provides high resolution spectra ($E$/$\Delta E$ up to 1000) in the 0.4-10~keV energy rage, taking advantage of both sets of gratings available within the HETG\footnote{See all details at \url{https://cxc.harvard.edu/proposer/POG/html/chap8.html}.}, i.e. the High Energy Grating (HEG) and the Medium Energy Grating (MEG). We excluded from our analysis the observation ID.~1199, as its exposure time is far too low ($\lesssim$2~ks) to perform any meaningful spectral investigation. The first 14 observations listed in the first half of Table~\ref{tab:obs} are those that were already reported by ISH06, while the following 13 observations were collected after the publication of their original paper and are reported here for the first time. These observations roughly double the exposure time available to perform the Doppler measurements and provide a larger orbital phase coverage (see Sect.~\ref{sec:discussion}). 

All data were processed with the {\sc chandra\_repro} tool (version 15 December 2022) available within the \chan\ data analysis software {\sc CIAO} v.4.14, exploiting also the most recently available (at the time of performing the analysis) calibration database (CALBD v.4.9.8). As Capella is a relatively bright X-ray source for the HETG, the {\sc chandra\_repro} script was run by setting the parameter {\sc tg\_zo\_position=detect} and we verified {\it a posteriori} that the {\sc tg\_findzo}  was correctly employed in all observations to correct pile-up issues and avoid a misposition of the source centroid peak in the zeroth-order image (ruling out poor zero-wavelength solutions). We also set the parameter {\sc pix\_adj=NONE} in order to be compliant with the previous approach adopted by ISH06 and avoid that a tighter point-spread-function, obtained as a consequence of the randomization, would lead to smaller uncertainties on the source localization and  impact on the outcomes of the {\sc tg\_findzo} processing.   

To obtain reliable measurements of the Doppler velocities of the emission lines in Capella's HETG spectra, we followed a similar technique as that described by ISH06. We briefly summarize this technique below, highlighting also the main differences between our and the original ISH06 approach. We first extracted for each \chan\ observation the $\pm1$st order MEG and HEG spectra (4 spectra per observation), building the corresponding arf and rmf matrices via the {\sc chandra\_repro} script. We inspected all spectra to verify that the previously identified emission line complexes used by ISH06 for the Doppler measurements (see their Table~2) were visible in all observations and determined the corresponding energy intervals in keV to perform the fits. As for ISH06, we considered a total of 6 energy intervals for the HEG spectra and 9 for the MEG (as the latter are extending the energy coverage down to 0.4~keV in a domain not available for the HEG spectra).  

Following ISH06 and other results in the literature \citep[see, e.g.,][and references therein]{yael}, we initially fit the entire MEG and HEG spectra in each order for every observation with a model comprising three collisionally-ionized diffuse gas components whose emission lines were both velocity- and thermally-broadened. Contrary to ISH06, we did not use the {\sc ISIS} environment for the spectral fitting but {\sc xspec} v.12.12.1 \citep{xspec} and the {\sc BAPEC}\footnote{For all details see \url{https://heasarc.gsfc.nasa.gov/xanadu/xspec/manual/node136.html}.} implementation for the gas components. In the fits to the entire spectra, the temperatures of the three components were fixed to $kT\sim10^{6.3}$~keV, $kT\sim10^{6.8}$~keV, and $kT\sim10^{7.1}$~keV, respectively. The metal abundances were fixed to solar values, while the Gaussian velocity broadening parameter, the redshift (providing the line velocity estimate), and the normalizations of the three components were left free to vary\footnote{Note that our goal here is to provide Doppler velocity measurements and we do not aim at achieving an exhaustive plasma modeling for Capella.}. 
The fits to the entire $\pm$1 order MEG and HEG spectra were used to determine the best values of the normalizations of the three {\sc bapec} components that were then fixed during the individual fits in the different energy-selected ranges with the prominent emission lines. During the fits to these restricted energy intervals, only the velocity broadening and the redshift parameters were left free to vary (linking together the parameters for the three components in each fit). We performed a total of 34 fits for each observation, counting also the 4 full spectra. In agreement with ISH06, we found that in virtually all fits the Gaussian velocity broadening parameter could not be significantly constrained, resulting always $<$300~km~s$^{-1}$ (at 90\% confidence level, hereafter c.l.). However, contrary to what has been reported by these authors, our analysis did not find any ``outlier'' in the measurements and thus we did not exclude the largest and lowest measured velocities in each observation. We retained all results for the energy intervals where the significance of the Doppler measurement was at $\sim3\sigma$~c.l. In the most recent observations with the lower exposure times, some of the emission lines at the softer energies ($\lesssim$1~keV) could be only marginally detected due to the degradation of the ACIS-S sensitivity \footnote{See https://cxc.cfa.harvard.edu/ciao/why/acisqecontamN0010.html}. These did not provide usable Doppler measurements. All redshift values measured from the $\sim$30 individual fits to the lines in each observation were then averaged to obtain the apparent Doppler measurement corresponding to that observation.  

All apparent Doppler measurements were then corrected for the barycentric motion of the Earth around the center of mass of the solar system. We used the algorithm developed by \citet{stumpff80} as implemented in the {\sc ispec} software \citep{ispec1,ispec2}. Following ISH06, we neglected the correction for the velocity of the spacecraft which is at least an order of magnitude lower than the barycentric correction and well within the uncertainties of all reported measurements. 

The outcomes of all this analysis are summarized in Table~\ref{tab:obs}. The uncertainties on the apparent Doppler measurements are reported at 3~$\sigma$ c.l. (as obtained from the fits in {\sc xspec}) to allow a direct comparison with ISH06. No systematic error has been added on top of the statistic uncertainties.  

We also recalculated the orbital phase corresponding to each \chan\ observation using the most updated Capella ephemerides provided by \citet{strassmeier20}. The start times of all \chan\ observations were first converted in heliocentric Julian day (HJD) and then scaled by using the refined orbital period reported by these authors to the epoch of passage at the ascending node (2454393.8621$\pm$1.0~HJD, private communication from M. Weber). The latter is assumed as phase 0 to ease the comparison with ISH06 results,  where the older ephemerides by \citet{hummel94} were adopted\footnote{Note that \citet{hummel94} assumed in their derivation of Capella's ephemerides a circular binary orbit and thus in their case the passage at the ascending node coincides with the periastron passage. The latest findings reported by \citet{strassmeier20} show that Capella's eccentricity is small but the orbit is not exactly circular. Therefore, in this case, the ascending node and the periastron passage are not occurring at the same epoch.}. 
\begin{figure*}
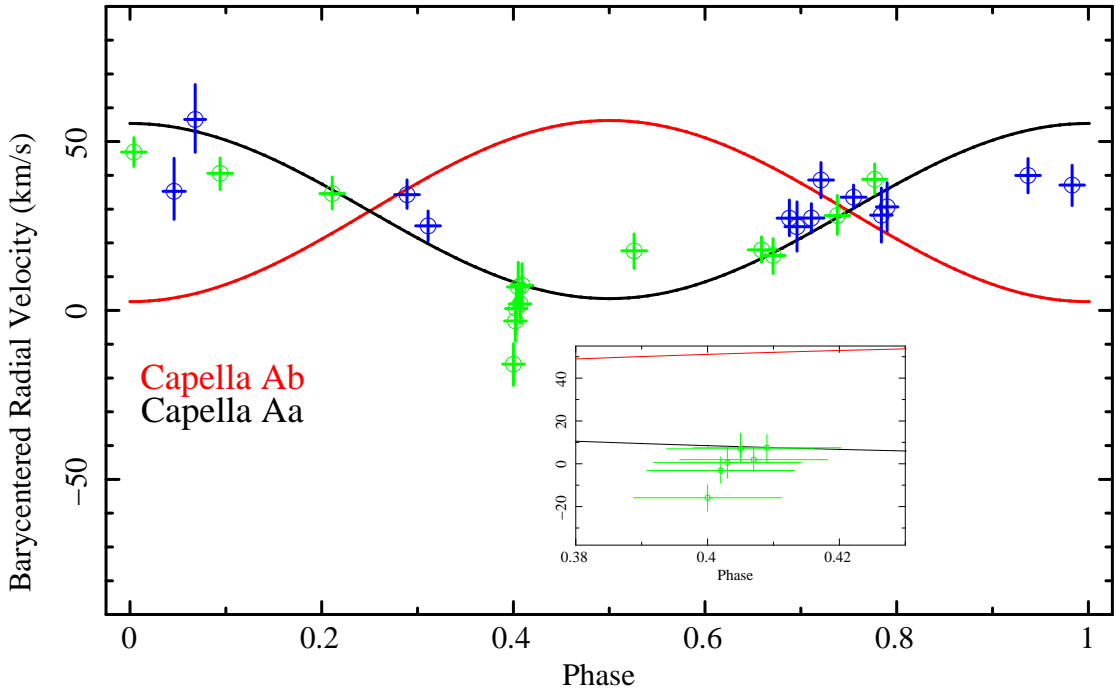

  \centering
  \includegraphics[width=9.0cm, angle=-90]{final_figure_1sigma.ps}
  \makebox[0pt][r]{
    \raisebox{-16.em}{
      \includegraphics[width=3.1cm, angle=-90]{insert_1sigma.ps}
    }\hspace*{9.5em}
  }%
  \caption{\label{results} Plot of the barycentered radial velocity measurements (green and blue points) obtained from all available \chan/HETG observations of Capella (see Table~\ref{tab:obs}). The blue points correspond to observations that were not reported before by ISH06. The error bars for each radial velocity measurement are given at 1~$\sigma$~c.l. and are only statistical. The uncertainty on the phase for each green point includes the duration of the observation, as well as the uncertainties associated to the ephemerides. The black and red lines are the expected barycentered radial velocities calculated according to the latest Capella's orbital parameters published by \citet{strassmeier20}. Uncertainties on these velocities were not included here as they are typically $\ll$1~km~s$^{-1}$. The phase 0 is assumed as the epoch of passage at the ascending node. The insert shows, for clarity, a zoom around the orbital phase 0.4 showing six observations closely overlapping in phase.} 
\end{figure*}

\section{Results and discussion}
\label{sec:discussion}

All results of our analysis are summarized in Fig.~\ref{results}. The black and red lines represent the barycentered radial velocity along the orbits of the two Capella's components (Aa and Ab) derived from the orbital parameters as in \citet{strassmeier20}. The green points with the corresponding error bars are the corrected Doppler velocities obtained from the \chan\ observations. When comparing our figure with Fig.~2 in ISH06, we note that there is a quantitative agreement between the results obtained for the observations analyzed in both works. The extension of the error bars is comparable (when compared at 3~$\sigma$ c.l.) and the estimated barycentered radial velocities are compatible to within these uncertainties. The 13 new observations reported here for the first time strengthen some of the previous conclusions reported by ISH06. First of all, by significantly extending the orbital phase coverage, the ensemble of the older and newer radial velocity measurements more firmly establish that Capella Aa is the dominant X-ray emitter in the binary, as the distribution of these velocities closely follow the predicted Doppler modulations expected for the coolest G8III giant (black line in Fig.~\ref{results}). Secondly, the observations providing the best signal-to-noise ratio (S/N) for the performed spectral analysis (i.e. the observations endowed with an exposure time $\gtrsim$20-30~ks) are indicating an overall uncertainty for the averaged barycentered radial velocity that is of other order of $\sim$20~km~s$^{-1}$ {(at 3~$\sigma$~c.l.)}. This was reported previously by ISH06 and was considered as a solid indication that the \chan/HETG is a stable instrument over the timescale of years. The extension of ISH06 results presented here further evidence the stability of the instrument (and the accuracy of the performed in-flight calibrations\footnote{See \url{https://cxc.harvard.edu/proposer/POG/html/chap8.html}.}) achieving now a baseline of over 22 years. 

A third important point of comparison between ISH06 results and those reported here concerns the measured barycentered radial velocities at phases 0-0.2. Contrary to the suggestion by ISH06, the usage of updated calibrations and the addition of two more data points at this orbital phase interval do not seem to show any longer the presence of a significant X-ray contribution to the emission lines by the secondary Capella Ab giant star (when uncertainties at 3~$\sigma$~c.l. are considered). Based on the results obtained from the observations ID.~5955 and 5040, ISH06 originally concluded that the somewhat lower radial velocity measurements they obtained could have been the result of contamination by the secondary’s (Capella Ab) line emission flux originating from its hottest 10$^7$~K plasma. Our revised measurement for the observation ID.~5955, although compatible to within the relatively large uncertainties with that reported by ISH06, is now in agreement with the expected value for the orbit of Capella Aa at this phase (at 3~$\sigma$~c.l.). The revised result for the observation ID.~5040 is still slightly lower than expected (according to the black line in Fig.~\ref{results}), but the two additional measurements now available within the same  orbital phase interval (ID.~21786 and 22003) do not provide any further indication for an overall lower systemic velocity  {(at 3~$\sigma$~c.l.)}. Unfortunately, both observations ID.~21786 and 22003 are characterized by a relatively low effective exposure ($\ll$20~ks) and the obtained radial velocity measurements are significantly more uncertain than those at other orbital phases. To further test the possibility that the \chan\ radial velocity measurements are affected by emission from the Ab giant, we performed a fit to the green points in Fig.~\ref{results} with the equation describing the black line and compared the results with a similar fit where the amplitude of the black line was left free to vary. The latter fit did not provide a significant  improvement over the former, returning an F-test probability of $\sim$0.11 (corresponding to less than 2~$\sigma$).  
The uncertainty estimated on the curve amplitude is compatible with the value corresponding to the plotted black line already at 90\% c.l. We thus conclude that, at present, the uncertainties on the measurements do not allow us to firmly exclude some possible contamination from the Ab giant, but there is no statistically significant indication of such contamination. In order to achieve a more decisive conclusion about the contamination, it would be advisable to schedule the future \chan/HETG calibration observations aimed at Capella during the orbital phases corresponding to the largest distances between the red and black lines in Fig.~\ref{results}, i.e. phases 0.0-0.2, 0.4-0.7, and 0.9-1.0. An exposure times of $\gtrsim$60~ks should be considered to improve the S/N of each measurement by at least a factor of $\sim$2 and reduce the currently available uncertainties. 

Based on the results that we can achieve after over 22 years of monitoring with the \chan/HETG, we learned that the centroid energies of Capella's most prominent X-ray emission lines produced by the coolest Aa component display a regular variability by up to $\sim$50~km~s$^{-1}$ at the most due to the stellar binarity of the system (see Fig.~\ref{results}). At present, it is reasonable to assume that there is not a significant contamination to these lines from the hotter Ab Capella component (and thus no measurable effect on the Doppler modulation of the lines' centroid energies over time). For the calibrations of the \chan/HETG, this effect was taken into account when measured a few years into the mission scientific operations \citep[][and see \url{https://cxc.harvard.edu/newsletters/news_12/hetg.html}]{ishibashi04}.

Concerning \xrism,\  the current plan is to use for the in-flight calibrations of Resolve mostly the so-called modulated X-ray sources installed on the instrument filter wheel \citep[MXSs;][]{mxs}. However, the limitation of the MXSs is that they only produce a few X-ray lines (mainly Cr-K and Cu-K lines at 5.41~keV and 8.04~keV, respectively) that do not homogeneously cover the instrument energy range. The same limitation applies to the radioactive $^{55}$Fe calibration sources that are installed both in one of the instrument filter wheel positions and above the so-called ``calibration'' pixel \citep[which is outside the field of view of the telescope and used only for calibration purposes;][]{xrismcal}. All these methods are sufficient to determine and monitor the detector electronic linear gain (see also Sect.~\ref{sec:intro}), but the identification and quantification of non-linear effects requires the observation of many emission lines across the entire instrument energy range and thus is best done by observing directly, e.g. coronal sources as Capella. The time variations of the emission lines in these objects affect the outcome of the calibrations and thus changes in energy driven, e.g., by stellar binarity have to be taken into account. The current Resolve in-flight calibration requirements aim at verifying an energy scale down to 2~eV and a full-width-half-maximum energy resolution of 1~eV for each pixel, adopting an in-flight calibration plan that also foresees to periodically observe Capella \citep{xrismcal}. Our work here provides the Resolve team a clear and updated reference where the time-dependent variability of Capella's emission lines velocities is quantified and can thus be corrected for.

\section*{Data availability}
All data exploited in this paper are publicly available from the \chan\ archive and processed with publicly available software.

\section*{Acknowledgements}
We thank the anonymous referee for useful comments. EB is grateful to Alessandro Papitto for useful discussions. Support for DPH was provided by NASA through the Smithsonian Astrophysical Observatory (SAO) contract SV3-73016 to MIT for Support of the Chandra X-Ray Center (CXC) and Science Instruments. CXC is operated by SAO for and on behalf of NASA under contract NAS8-03060. 

\bibliography{bib.bib}{}
\bibliographystyle{mnras}

\end{document}